\documentclass[sigconf]{acmart}

\usepackage{graphicx}
\usepackage{amsfonts}
\usepackage{amsmath}
\usepackage{algorithm}
\usepackage[noend]{algpseudocode}
\usepackage[utf8]{inputenc}
\usepackage{circledsteps}
\usepackage{comment}
\usepackage[most]{tcolorbox}
\usepackage{multirow}
\usepackage{tikz}
\usepackage{circledsteps}
\usepackage{lscape}
\usepackage{xcolor}
\usepackage{colortbl}
\usepackage{booktabs}
\usepackage{tabularx}
\usepackage{array}
\usepackage{url}
\usepackage{hyperref}
\usepackage{balance}
\usepackage{longtable} 

\AtBeginDocument{%
  \providecommand\BibTeX{{%
    \normalfont B\kern-0.5em{\scshape i\kern-0.25em b}\kern-0.8em\TeX}}}

\setcopyright{acmlicensed}
\copyrightyear{2025}
\acmYear{2025}
\acmDOI{XXXXXXX.XXXXXXX}

\acmConference[EASE 2025]{The 29th International Conference on Evaluation and Assessment in Software Engineering}{17–20 June, 2025}{Istanbul, Türkiye}
\acmISBN{978-1-4503-XXXX-X/18/06}

\begin{document}

\title{Racing Against the Clock: Exploring the Impact of Scheduled Deadlines on Technical Debt}



\author{Joshua Aldrich Edbert}
\email{joshua.edbert@usask.ca}
\affiliation{%
  \institution{University of Saskatchewan}
  \city{Saskatoon}
  \country{Canada}
}

\author{Zadia Codabux}
\email{zadiacodabux@ieee.org}
\affiliation{%
  \institution{University of Saskatchewan}
  \city{Saskatoon}
  \country{Canada}
}

\author{Roberto Verdecchia}
\email{roberto.verdecchia@unifi.it}
\affiliation{%
  \institution{University of Florence}
  \city{Florence}
  \country{Italy}
}


\begin{abstract}

\textbf{Background:} Technical Debt (TD) describes suboptimal software development practices with long-term consequences, such as defects and vulnerabilities. Deadlines are a leading cause of the emergence of TD in software systems. While multiple aspects of TD have been studied, the empirical research findings on the impact of deadlines are still inconclusive. \textbf{Aims:} This study investigates the impact of scheduled deadlines on TD. It analyzes how scheduled deadlines affect code quality, commit activities, and issues in issue-tracking systems. \textbf{Method:}  We analyzed eight Open Source Software (OSS) projects with regular release schedules using SonarQube. We analyzed 12.3k commits and 371 releases across these eight OSS projects. The study combined quantitative metrics with qualitative analyses to comprehensively understand TD accumulation under scheduled deadlines. \textbf{Results:} Our findings indicated that some projects had a clear increase in TD as deadlines approached (with above 50\% of releases having increasing TD accumulation as deadlines approached), while others managed to maintain roughly the same amount of TD. Analysis of commit activities and issue tracking revealed that deadline proximity could lead to increased commit frequency and bug-related issue creation. \textbf{Conclusions:} Our study highlights that, in some cases, impending deadlines have a clear impact on TD. The findings pinpoint the need to mitigate last-minute coding rushes and the risks associated with deadline-driven TD accumulation.

\end{abstract}




\maketitle

\section{Introduction}

Technical Debt (TD), first introduced by Ward Cunningham in 1992~\cite{cunningham1992wycash}, refers to the non-ideal outcomes stemming from particular design and implementation decisions made for short-term gains that may lead to complications in the maintenance and future development of a software project~\cite{kruchten2012technical}. While TD can arise at any stage of the software development process, it predominantly affects the maintenance phase~\cite{alves2016identification}. If not properly managed, TD can lead to escalated costs and diminished product quality, impeding the sustained progress of software development efforts~\cite{lim2012balancing}. Consequently, TD is recognized as a significant issue within the field of software engineering~\cite{tom2013exploration}.

There has been a notable focus on managing scheduling and temporal concerns within software engineering~\cite{kuutila2020time}. Deadlines in software development scheduling are a top cause of the emergence of quality issues and TD in software systems~\cite{almeida2023behind, ramavc2022prevalence, verdecchia2021building}. Schedule constraints are often linked to highly detrimental consequences, including the creation of a chaotic work environment, high-stress levels for developers, and the emergence of stress-induced health problems, ultimately decreasing team productivity and effectiveness~\cite{borg2019video, kuhrmann2016teams}. 

Despite deadlines being one of the top reasons developers accumulate TD in their software projects, no previous studies have analyzed how deadlines affect TD in software projects with regular release schedules. The empirical research findings on the impact of deadlines are still inconclusive~\cite{basten2021impact}. Investigating TD can enhance internal software quality attributes by detecting recurring code violations~\cite{johnson2013don}. For instance, by examining the patterns of TD values for each commit,  alongside other software metrics, such as commit frequency or the bug types recorded in the issue tracking system, we can identify the specific factors contributing to recurring code violations, allowing for a clearer understanding how scheduled deadlines cause TD within the development process. This analysis can also reveal whether TD accumulates steadily over time or intensifies at certain points, such as near impending project deadlines, helping teams monitor and predict high-risk periods for quality issues. Additionally, examining the impact of deadlines in software development can help maintain a balanced workload for developers, prevent burnout, and consequently enhance the performance of each team member~\cite{kuutila2017reviewing}. Hence, this research contributes to the broader discourse of deadlines in software development, specifically on TD. Our study provides concrete empirical evidence on how scheduled deadlines impact TD in software projects.  

This paper investigates the impact of scheduled deadlines on TD in Open Source Software (OSS) projects. The study analyzed projects using SonarQube\footnote{sonarsource.com} to assess the code quality regarding TD. This approach provides a quantifiable measure of the TD accumulated over the lifecycle of a project and how it fluctuates concerning regular release deadlines. The research also examined the influence of impending project deadlines on development commit activities and issues created. Information such as code churn, commit frequency, issue creation and closing date, and issue labels were scrutinized to discern patterns that may emerge as deadlines approached. The scope of our analysis in this study is solely on code debt, a specific category of TD~\cite{li2015systematic}. We acknowledge the existence of various types of TD, such as architectural or design debt. We focused only on code debt, as it is the most studied TD type~\cite{li2015systematic}. This focus allows us to delve deeper into the nuances of code quality issues under scheduled deadlines, providing a more detailed exploration within a constrained scope. Note that code debt and TD are used interchangeably in this paper. 

This study is the first to describe the impact of scheduled deadlines on TD accumulation, contributing to a comprehensive understanding of how scheduling influences code quality. Our key contributions are:
\begin{enumerate}
    \item We provide the first concrete empirical data on how scheduled deadlines impact TD accumulation in software projects, filling a critical gap in the existing literature. This information enables teams to anticipate and mitigate high-risk periods for quality issues.

    \item Our study offered insights into software project commit activities under deadlines to better understand how development teams adjust their workflow as deadlines approach. This insight helps identify operational patterns that could lead to an increase in TD, providing a foundation for more strategic planning in software projects.

    \item We improved the understanding of software project issue management under scheduled deadlines. This investigation reflects the immediate responses of development teams to deadlines and how such conditions may rush or neglect the resolution of critical issues, contributing to the accumulation of TD.
    
    \item A comprehensive replication package\footnote{\url{https://zenodo.org/records/14782835}} of the study.
\end{enumerate}


\section{Related Work}
\label{sec:rw}

\textbf{Technical Debt Evolution Studies.} Maggi et al.~\cite{maggi2025evolution} explored how code debt changed over time within 13 microservice-based projects. Their study used SonarQube to analyze TD, applied statistical methods to detect trends in TD, and followed up with a manual inspection of commits. They found that TD generally grew over time, although periods of stability did occur. Molnar et al.~\cite{molnar2020long} analyzed the evolution and characteristics of TD in three Java-based open-source applications. Using SonarQube for a detailed analysis of 110 versions, the study identified crucial versions pivotal in accumulating or reducing TD. Openja et al.~\cite{openja2022technical} explored the distribution and evolution of TD in quantum software and how they correlated with the emergence of faults. They conducted an empirical study of 118 open-source quantum projects sourced from GitHub, organized into ten distinct categories. They found that quantum software projects commonly face challenges related to code convention breaches, error handling issues, and design flaws. Lenarduzzi et al.~\cite{lenarduzzi2020does} studied how shifting from a traditional monolithic system to a microservices framework affected TD. The analysis involved a combination of TD measurement using SonarQube and interviews with company members. The findings revealed that although initially, there was an increase in TD following the introduction of microservices, this trend reversed over time. Digkas et al.~\cite{digkas2017evolution} analyzed sixty-six Java OSS projects within the Apache ecosystem, examining the evolution of TD. Their results indicated an overall increase in TD and source code metrics across most systems. When normalized to the system's size, TD declined over time in most cases. Mamun et al.~\cite{al2019evolution} delved into TD evolution, employing a novel ``technical debt density trend" metric to assess debt changes against code growth. Analyzing 21 OSS projects via SonarQube revealed that TD initially grew with smaller code volumes but lessened as the code expanded. 

\textbf{Time-Constraints in Software Engineering Studies}. Austin \cite{austin2001effects} utilized an agency framework to explore the impact of time pressure on software development quality, focusing on developers' behavior under deadline constraints. It analyzed the trade-off between maintaining product quality and meeting deadlines, suggesting that developers may resort to shortcuts that compromise quality. The research findings suggested that aggressive deadline-setting might prevent shortcut-taking by maintaining high time pressures. Basten et al.~\cite{basten2021impact} evaluated the impact of time pressure on software quality through two experiments, testing a game-theoretical model that posits high levels of time pressure can deter developers from taking shortcuts, thus improving software quality. The first experiment involved abstract decision-making with financial trade-offs, while the second focused on actual programming tasks under varied deadline conditions. The findings supported the model, showing that higher probabilities of unrealistic deadlines led to higher software quality. M{\"a}ntyl{\"a} et al.~\cite{mantyla2014time} conducted a controlled experiment to investigate the effects of time pressure on the efficiency and effectiveness of software engineering tasks, specifically test case development and requirements review. They utilized 97 observations from 54 subjects. Results indicated a significant increase in efficiency under time pressure without a corresponding decrease in effectiveness or negative impacts on motivation or frustration. The study suggested that moderate time pressure could enhance productivity in well-defined software engineering tasks. Salman et al.~\cite{salman2019controlled} explored how time constraints affected confirmation bias in functional software testing. Forty-two graduate students were asked to create functional test cases with or without time pressure. The findings revealed that although participants tended to create test cases that confirmed their beliefs, time pressure did not notably increase this tendency. Shah et al.~\cite{shah2014global} investigated outsourced offshored global software testing practices to understand its unique challenges. Over two months, researchers used ethnographic methods, including interviews and observations. Key findings showed that test engineers' motivation and recognition significantly impacted testing quality. The study also found that pressures increased with intermediate onshore teams. Malgonde et al.~\cite{malgonde2014applying} explored the roots and impacts of time pressure on agile methods and used Extreme Programming as a case study for its adaptive strategies. Interviews with project managers about handling increased time pressures aim to validate these controls' effectiveness in agile environments, offering insights into adapting agile practices under tight timelines. Fehrenbacher et al.~\cite{fehrenbacher2014behavioural} examined the effects of time pressure and justification requirements on software acquisition decision-making. It used eye-tracking to study how these factors influenced cognitive and behavioral responses. The findings revealed that time pressure increased discomfort and reduced time spent on information analysis. Importantly, requiring justification under time pressure led to a more thorough examination of information. Lohan et al.~\cite{lohan2014investigation} explored how different types of time pressure and the cohesiveness of a group influenced the quality of decisions made in software development teams. The authors surveyed 119 software developers to assess the effects of challenge-based and hindrance-based time pressures on the quality of decision-making, measured by the confidence and consensus of teams in their decisions. The results indicated that decision-making quality was positively influenced by challenge time pressures and the cohesion of the group, while hindrance time pressure had no notable effect.

\textbf{Technical Debt Management Studies}. Tan et al.~\cite{tan2023lifecycle} investigated the management of TD across issue trackers and software repositories, aiming to understand the lifecycle from TD introduction to resolution. They analyzed 3000 issues from five projects, focusing on 300 issues with TD and tracking 312 TD items. Key findings included that TD identification often took about a year, but resolution times were shorter when the same developer was involved throughout the TD lifecycle. Chatzigeorgiou et al.~\cite{chatzigeorgiou2010investigating} aimed to investigate the development of bad smells or software design problems in object-oriented programming. They also sought to understand whether bad smells were resolved naturally over time or required specific intervention. The study analyzed historical code versions from two OSS systems, focusing on three types of bad smells. Tan et al.~\cite{tan2021evolution} analyzed TD remediation in 44 Python projects from the Apache Software Foundation, focusing on the types and volume of debt addressed. Analysis of extensive data showed that most efforts were made to target testing, documentation, complexity, and duplication issues, with over half of the TD being short-term and resolved within two months. Digkas et al.~\cite{digkas2018developers} analyzed how TD was managed in fifty-seven Java OSS projects under the Apache Software Foundation, mainly focusing on the issues addressed and the amount of debt repaid. The study discovered that only a few issues accounted for most TD repayments. It explored several aspects of TD management, including differences in issue resolution rates among projects, the frequency of fixes for various issue types, the distribution of effort in addressing different types of debt, and the timeframe for debt repayment. They found a substantial portion of issues, nearly 20\%, are fixed within a month of detection, with more than half resolved within a year.

\textbf{Summary:} Unlike most prior research that explored time constraints or software quality independently, our study took a distinct approach to empirically investigate the interplay between deadlines and software quality, focusing on TD. We applied ASAT to obtain quantifiable, time-based measures of TD accumulation. These analyses allowed us to observe how software quality evolved regarding scheduled deadlines based on software development activities. No previous study has analyzed how TD was accumulated with regard to scheduled deadlines using empirical data, nor have trend-based analyses been conducted to examine the temporal dynamics of TD around release deadlines.


\section{Methodology}
\label{sec:methodology}
This section outlines the study methodology, detailing the goal, research questions, data collection, and analysis. 

\subsection{Goal and Research Questions}
The aim of this study is (\textcolor{black}{articulated using the Goal-Question-Metric approach~\cite{gqm}}):
\begin{quote}
\textbf{Purpose:} To investigate\\
\textbf{Issue:} the impact\\
\textbf{Object:} of scheduled deadlines on technical debt\\
\textbf{Viewpoint:} from the software engineering researchers' perspective
\end{quote}

\textbf{$\mathbf{RQ_1}$ What is the impact of scheduled deadlines on TD accumulation?}

\textit{Rationale: } The rationale behind investigating the impact of scheduled deadlines on TD accumulation lies in understanding how regular deadlines compel teams to take shortcuts or rush their work, potentially increasing TD. This insight is vital as it provides a data-driven perspective of the risks associated with deadlines for TD management. Furthermore, it contributes to the broader body of software engineering knowledge by providing empirical evidence on the dynamics of TD and scheduled deadlines.

\textbf{$\mathbf{RQ_2}$ What is the impact of scheduled deadlines on the development commit activities?}

\textit{Rationale: }The investigation into the impact of scheduled deadlines on development activities aims to shed light on developer behavior under deadlines, revealing whether changing code churn, commit frequency, or file changes occur as deadlines approach. The insights could reveal whether teams tend to work more intensively and make more changes as regular deadlines approach~\cite{camuto2021suite, alali2008s}. Understanding these trends could help plan future projects more effectively. If specific patterns consistently led to increased TD, they could be addressed early in the project lifecycle.

\textbf{$\mathbf{RQ_3}$ What is the impact of scheduled deadlines on the project issues?}

\textit{Rationale: }In modern software development projects, issue trackers have emerged as vital tools for collaboration~\cite{bertram2010communication}. Issue trackers monitor new feature requests, development tasks, and bugs~\cite{kikas2015issue}. By analyzing patterns in issue creation, closing dates, and issue labels, our study aims to understand how scheduled deadlines influence issue management in software projects. This aspect is crucial as it directly reflects how teams respond to and resolve issues under scheduled deadlines. Patterns in issue management could reveal whether deadlines led to a hurried resolution of issues, potentially leaving some issues inadequately addressed or ignored, thus contributing to the accumulation of TD.

\vspace{-5px}
\subsection{Data Collection}
\subsubsection{Project Selection}
\label{Section:ProjectSelection}

\begin{table*}[ht]
\centering
\caption{The OSS Projects Selected in this Study}
\label{datacollectiontable}
\begin{tabularx}{\linewidth}{l|X|X|X|X|X|X|X|X}
\toprule
\textbf{} & \textbf{Home Assistant} & \textbf{GitLab} & \textbf{Django} & \textbf{Kubernetes} & \textbf{Jenkins} & \textbf{Eclipse} & \textbf{Go} & \textbf{Ansible} \\
\midrule
\multicolumn{1}{l|}{Project Size (LOC)} & \multicolumn{1}{r|}{2,348,208} & \multicolumn{1}{r|}{9,098,312} & \multicolumn{1}{r|}{656,302} & \multicolumn{1}{r|}{4,645,775} & \multicolumn{1}{r|}{342,060} & \multicolumn{1}{r|}{802,165} & \multicolumn{1}{r|}{2,255,066} & \multicolumn{1}{r}{240,188} \\
\multicolumn{1}{l|}{Project Age (Months)} & \multicolumn{1}{r|}{137} & \multicolumn{1}{r|}{161} & \multicolumn{1}{r|}{237} & \multicolumn{1}{r|}{129} & \multicolumn{1}{r|}{221} & \multicolumn{1}{r|}{288} & \multicolumn{1}{r|}{639} & \multicolumn{1}{r}{156} \\
\multicolumn{1}{l|}{Contributors (People)} & \multicolumn{1}{r|}{4,021} & \multicolumn{1}{r|}{6,976} & \multicolumn{1}{r|}{2,614} & \multicolumn{1}{r|}{3,757} & \multicolumn{1}{r|}{795} & \multicolumn{1}{r|}{180} & \multicolumn{1}{r|}{2,168} & \multicolumn{1}{r}{5,566} \\
\multicolumn{1}{l|}{Stars, Forks} & \multicolumn{1}{r|}{75.6k, 31.9k} & \multicolumn{1}{r|}{5.3k, -} & \multicolumn{1}{r|}{81.8k, 32k} & \multicolumn{1}{r|}{112k, 40k} & \multicolumn{1}{r|}{23.5k, 8.9k} & \multicolumn{1}{r|}{88, 115} & \multicolumn{1}{r|}{125k, 17.8k} & \multicolumn{1}{r}{63.6k, 23.9k} \\
\multicolumn{1}{l|}{Classes} & \multicolumn{1}{r|}{12,821} & \multicolumn{1}{r|}{35,030} & \multicolumn{1}{r|}{12,919} & \multicolumn{1}{r|}{20,204} & \multicolumn{1}{r|}{8,587} & \multicolumn{1}{r|}{10,904} & \multicolumn{1}{r|}{14,003} & \multicolumn{1}{r}{2,112} \\
\multicolumn{1}{l|}{Repository Commits} & \multicolumn{1}{r|}{120,451} & \multicolumn{1}{r|}{434,834} & \multicolumn{1}{r|}{33,153} & \multicolumn{1}{r|}{127,317} & \multicolumn{1}{r|}{35,713} & \multicolumn{1}{r|}{44,263} & \multicolumn{1}{r|}{61,914} & \multicolumn{1}{r}{54,746} \\
\cmidrule{1-9}
\multicolumn{1}{l|}{Language} & Python & Ruby, Go, JavaScript & Python & Go, Shell, PowerShell & Java & Java & Go & Python \\
\cmidrule{1-9}
\multicolumn{1}{l|}{Domain} & Home Automation & Version Control System & Web Development & Container & Server Automation & IDE & Language & IT Automation Platform \\
\cmidrule{1-9}
\multicolumn{1}{l|}{Release Schedule} & First Wednesday of each month & 22nd of each month & Every month & Three times a year & Every Tuesday & December, March, June, September & August and February & November and May \\
\bottomrule
\end{tabularx}
\end{table*}

The initial step of our study is cloning OSS projects with regularly scheduled releases. We followed several criteria to maximize the variety and representativeness of our selected projects and reduce potential external validity threats, as established by previously published studies~\cite{lenarduzzi2019technical, codabux2020profiling}. Specifically, (1) we considered Patton's ``criterion sampling" approach~\cite{patton1990qualitative}, selecting projects that were older than four years, had more than 500 commits, contained over 100 classes, and recorded more than 100 issues in their issue tracking system, (2) following the guidance of Nagappan et al.~\cite{nagappan2013diversity}, we ensured diversity in the selection by considering projects of varying ages, sizes, and domains. Moreover, we applied the following Selection Criteria (SC) from previous work~\cite{verdecchia2023tracing}: \textbf{SC1:} The use of real applications. This criterion helps exclude toy projects and demos. 
\textbf{SC2:} The number of times the repository is forked and starred. This criterion provides assurance about the quality and popularity of the repository. \textbf{SC3:} The number of commits to the repository. This ensures the project is representative of a long-lived application. Our selection process resulted in eight projects, namely Home Assistant, Django, Kubernetes, Jenkins, Eclipse, Go,  Ansible, and  GitLab, as summarized in Table \ref{datacollectiontable}. Notably, we selected eight OSS projects, which is more than the number of projects typically considered in similar studies~\cite{molnar2020long, codabux2020profiling}, underscoring the comprehensiveness of our approach. 

\subsubsection{SonarQube Analysis (RQ1)}
We focused on understanding the impact of scheduled deadlines on TD accumulation. For each of the selected projects, a daily snapshot of the project codebase was taken, specifically analyzing the last commit of each day. We utilized the time of the last commit on a given day as a purposive sampling strategy~\cite{marshall1996sampling} to limit the number of samples to be analyzed to a feasible magnitude while allowing us to gain a systematic overview of how TD varied in time. We considered the last commit of each day for the SonarQube analyses, which allowed us to study the evolution of TD throughout a wide temporal range (e.g., years) while maintaining the number of data points to be collected at a feasible value. This constitutes a tradeoff between construct validity and the feasibility of this study. Section \ref{sec:threats} further discusses this limitation of our study.

We analyzed 12.3k commits and 371 releases across the eight selected OSS projects. SonarQube, a tool for static code analysis, was used to detect TD induced by the developers on the daily snapshot of the projects. SonarQube was chosen as it identifies and measures TD and has been widely used in various studies~\cite{alfayez2023sonarqube, alfayez2018exploratory, verdecchia2024technical, molnar2020long, lenarduzzi2020does}. Furthermore, the TD metrics provided by SonarQube have been the most frequently employed in empirical evaluations within academic research~\cite{avgeriou2020overview}. For each snapshot, SonarQube version 9.9 LTS and SonarScanner\footnote{\url{docs.sonarsource.com/sonarqube/9.9/analyzing-source-code/scanners/sonarscanner/} Accessed on May 6, 2024} for Linux version 5.0.1 were used to examine every change. This procedure gathered the SQALE index metric~\cite{letouzey2012sqale}. The SQALE\footnote{docs.sonarsource.com/sonarqube/latest/user-guide/metric-definitions/} index is defined as \textit{``A measure of effort to fix all code smells. The measure is stored as minutes"}. To prevent subjective tool setting tempering, the standard SonarQube rules setup was utilized for TD measurement, as suggested by a previous study~\cite{verdecchia2023tracing}.

\subsubsection{Commit Activity (RQ2)}
The second research question aimed to examine the impact of scheduled deadlines on the development of commit activities, such as the commit frequency of the selected projects' developers. Pydriller~\cite{spadini2018pydriller} is a Python framework designed for mining information from Git repositories, commonly used in previous studies~\cite {palit2023automatic, bhandari2021cvefixes, camuto2021suite, radu2019dataset}. It facilitates the retrieval of data from a Git repository, including commit frequency, file change count, and code churn. We used Pydriller to analyze the eight selected projects over their lifespan, extracting metrics such as commit frequency, files changed, and code churn. Each day's commits were traversed to calculate these metrics for each project. The results captured aggregated daily summaries of the metrics. Our approach provided a detailed temporal view of repository activity, enabling insights into patterns like increased activity near deadlines or consistent development trends.

\subsubsection{Issue Tracking System (RQ3)}
A systematic analysis of the project issue-tracking system was conducted to examine the impact of scheduled deadlines on the issues arising. Our focus was on extracting comprehensive metadata associated with each issue, which provides insights into how the frequency, severity, or nature of these issues might change as deadlines approached. We began our data collection by interfacing with the project issue-tracking system, ensuring access to all the relevant issues since the inception of the project. We utilized the APIs provided by GitHub, GitLab, and Jira, which allowed for an automated and efficient extraction of issue data for all our selected projects, excluding Django. For Django, they made their issue tracker data readily downloadable as a CSV file on their website. For each issue, we extracted the following metadata: the unique issue number, the timestamp when the issue was created, its closure, and the associated labels. 

\subsection{Data Analysis}
\label{Section:DataAnalysis}

\begin{table*}[h!]
\centering
\caption{Taxonomy of the SQALE Index Derivatives Evolution Trends}
\label{tab:sonarqube_window_taxonomy}
\small
\begin{tabular}{>{\bfseries}l p{0.7\textwidth}}
\toprule
\textbf{Trend Category} & \textbf{Description} \\
\midrule
\addlinespace
Increase (I) & The trend generally shows an upward direction, although it may include brief periods of decline. \\
\addlinespace
Increasing Plateau (IP) & The density levels out at plateau(s) after an increasing trend that generally does not decline. \\
\addlinespace
Plateau Increasing (PI) & The trend initially remains steady and then exhibits growth over time. \\
\addlinespace
Decrease (D) & This is the opposite of \textit{I}, showing a general downward trend. \\
\addlinespace
Decreasing Plateau (DP) & This trend is the inverse of \textit{IP}, where the trend eventually stabilizes after showing an initial decreasing trend. \\
\addlinespace
Plateau Decreasing (PD) & This trend starts steadily and then shows a decline throughout the observation, serving as the opposite of \textit{PI}. \\
\addlinespace
Hill (H) & The trend starts at lower levels, ascends to a peak, and then descends, resembling the shape of a hill. \\
\addlinespace
Valley (V) & This is the counterpart to \textit{H}, where the trend dips to a low point and then rises, forming a valley-like shape. \\
\addlinespace
Valley Hill (VH) & The initial trend descends, reaching a bottom dip, before it ascends to achieve a peak, and then it descends again, resembling a valley followed by a hill. \\
\addlinespace
Hill Valley (HV) & The trend begins with an ascent to a peak, then a descent to a low point, and rises again, mimicking a hill followed by a valley. \\
\addlinespace
Double Hills (HH) & The density exhibits two peaks separated by a valley, indicating a pattern of increase to a peak, decrease, and then another increase to a second peak. \\
\addlinespace
Constant (C) & The trend line remains flat over time, indicating no change in the measured density. \\
\addlinespace
Anomalous (A) & This category is reserved for trends that do not fit into any of the predefined patterns above. \\
\bottomrule
\end{tabular}
\end{table*}

We analyzed the derivatives of the SQALE index values, which indicated the rate at which the SQALE index value increased or decreased daily. We extracted the SonarQube SQALE index values of the project at the last commit of each day. Our approach to analyzing the derivatives of the SQALE index values indicates how much the SQALE index value, a measure of effort (in minutes) to fix all code smells, is accumulated as deadlines approach instead of the SQALE index value, providing a nuanced understanding of the dynamics of TD accumulation in relation to scheduled deadlines. We acknowledged the potential single-metric bias of relying solely on the SQALE index for TD evaluation. Another limitation is the analysis of the derivatives of the SQALE index values without normalizing them against a size-related metric, such as LOC. Our decision was deliberately made, as Graf-Vlachy and Wagner~\cite{graf2023type} indicated that ratios could be problematic when used as dependent variables~\cite{certo2020divided}. Both threats will be further discussed in Section \ref{sec:threats}.

We plotted the SQALE index derivative trends for all projects as a time series for visualization and analysis. However, we only included a few plots in this paper due to space limitations, with the remaining data readily available in the replication package. Figure~\ref{fig:rq1} presents the trend for only one project as an example. This figure illustrates the SQALE index derivative trends over time for the Home Assistant project during 2022, providing a clear view of the temporal evolution of TD. The red dashed vertical lines in the plot represent the scheduled release deadlines, offering a contextual timeline to examine how TD accumulates and evolves around these scheduled deadlines. We qualitatively analyzed the temporal trends of the SQALE index derivative values across different projects, following the time-windowing guidelines from previous work~\cite{malavolta2018maintainability}. We defined a ``window" as the interval between two consecutive deadlines within a project, during which we observed and analyzed the evolution of the derivatives of the SQALE index values to determine the TD accumulation trends. To systematically categorize and understand these trends in each window, we employed the provisional coding technique~\cite{saldana2021coding} from the trends reported by Malavolta et al.~\cite{malavolta2018maintainability}, whose work also involved manually evaluating trends. We used provisional coding to develop an organized classification of the SQALE index derivative trends, enabling us to discern patterns indicating how TD accumulation evolved in relation to project deadlines. Table \ref{tab:sonarqube_window_taxonomy} defines the TD accumulation evolution pattern we used in our manual window characterization. Initially, we adopted the trend definitions provided by Malavolta et al.~\cite{malavolta2018maintainability} as the foundation for our categorization protocol, given their relevance and similarity to our objectives. Subsequently, we expanded on these definitions by introducing additional trend categories to capture the nuances observed in our data better. One author categorized the derivatives of the SQALE index trend per window for each project. Another author examined the categorization by the other to verify the correctness of the coding process. Any discrepancies were jointly discussed. This process was repeated for all windows across all projects to ensure comprehensive coverage. We obtained a very strong agreement between the two authors (Cohen’s Kappa = 0.936). 

Our decision to adopt provisional coding, rather than relying on statistical tests, was intentional and aligned with the goal of this study. First, we focused on providing qualitative insights into TD evolution. While valuable for identifying correlations, traditional statistical approaches lack the contextual granularity needed to understand the nuanced patterns present in our data. Moreover, the shape and trajectory of TD accumulation trends, such as \textit{Double Hills}, carry interpretive meaning that statistical coefficients alone cannot convey. The same method was applied in earlier research~\cite{malavolta2018maintainability}, further reinforcing our confidence in its use and suitability for our study.

RQ2 and RQ3 metrics were aggregated and analyzed over defined periods surrounding each deadline. Specifically, we examined whether notable spikes in the metrics' data occurred \textit{Exactly on the Deadlines (Ex), Right Before (Bf), Right After (Af),} or \textit{No Spikes at All Around the Deadlines (Ot)}. For the \textit{Bf} and  \textit{Af} categories, we used a 7-day window on either side of the deadline, inspired by a previous study~\cite{balasubramanian2018deadlines}.

\vspace{-5px}
\section{Results}
\label{sec:results}
\subsection{$\mathbf{RQ_1}$: The Impact of Scheduled Deadlines on TD Accumulation }


\begin{table*}[htbp]
\caption{Summary of RQ1 Results}
\centering
\begin{tabularx}{\textwidth}{l*{14}{>{\raggedleft\arraybackslash}X}}
\toprule
\textbf{Projects} & \textbf{V} & \textbf{H} & \textbf{VH} & \textbf{HV} & \textbf{HH} & \textbf{I} & \textbf{IP} & \textbf{PI} & \textbf{D} & \textbf{DP} & \textbf{PD} & \textbf{C} & \textbf{A} \\
\midrule
\textcolor{black}{HomeAsst} & \textbf{56.3\%} & 9.4\% & 0 & 0 & 0 & 12.5\% & 0 & 0 & 3.1\% & 3.1\% & 0 & 0 & 15.6\% \\
GitLab & \textbf{50.0\%} & 22.7\% & 4.5\% & 2.3\% & 0 & 9.1\% & 0 & 0 & 2.3\% & 0 & 0 & 0 & 9.1\% \\
Django & 20.6\% & \textbf{42.6\%} & 1.5\% & 1.5\% & 0 & 19.1\% & 1.5\% & 0 & 8.8\% & 0 & 0 & 2.9\% & 1.5\% \\
Kubernetes & 7.4\% & \textbf{74.1\%} & 3.7\% & 3.7\% & 3.7\% & 0 & 0 & 0 & 0 & 0 & 0 & 7.4\% & 0 \\
Jenkins & 12.0\% & 5.4\% & 0 & 0 & 0 & 17.4\% & 0.7\% & 0.7\% & 16.8\% & 0 & 2.0\% & \textbf{43.0\%} & 2.0\% \\
Eclipse & 8.0\% & 8.0\% & 4.0\% & 0 & 8.0\% & 16.0\% & 0 & 0 & 0 & 0 & 4.0\% & 4.0\% & \textbf{48.0\%} \\
Go & 5.6\% & 22.2\% & 0 & 0 & \textbf{27.8\%} & 5.6\% & 5.6\% & 11.1\% & 0 & 0 & 11.1\% & 0 & 11.1\% \\
Ansible & \textbf{50.0\%} & 16.7\% & 33.3\% & 0 & 0 & 0 & 0 & 0 & 0 & 0 & 0 & 0 & 0 \\
\bottomrule
\end{tabularx}
\label{tab:rq1}
\end{table*}

\begin{figure}[!ht]
\centering
\includegraphics[width=8.5cm, height=4cm]{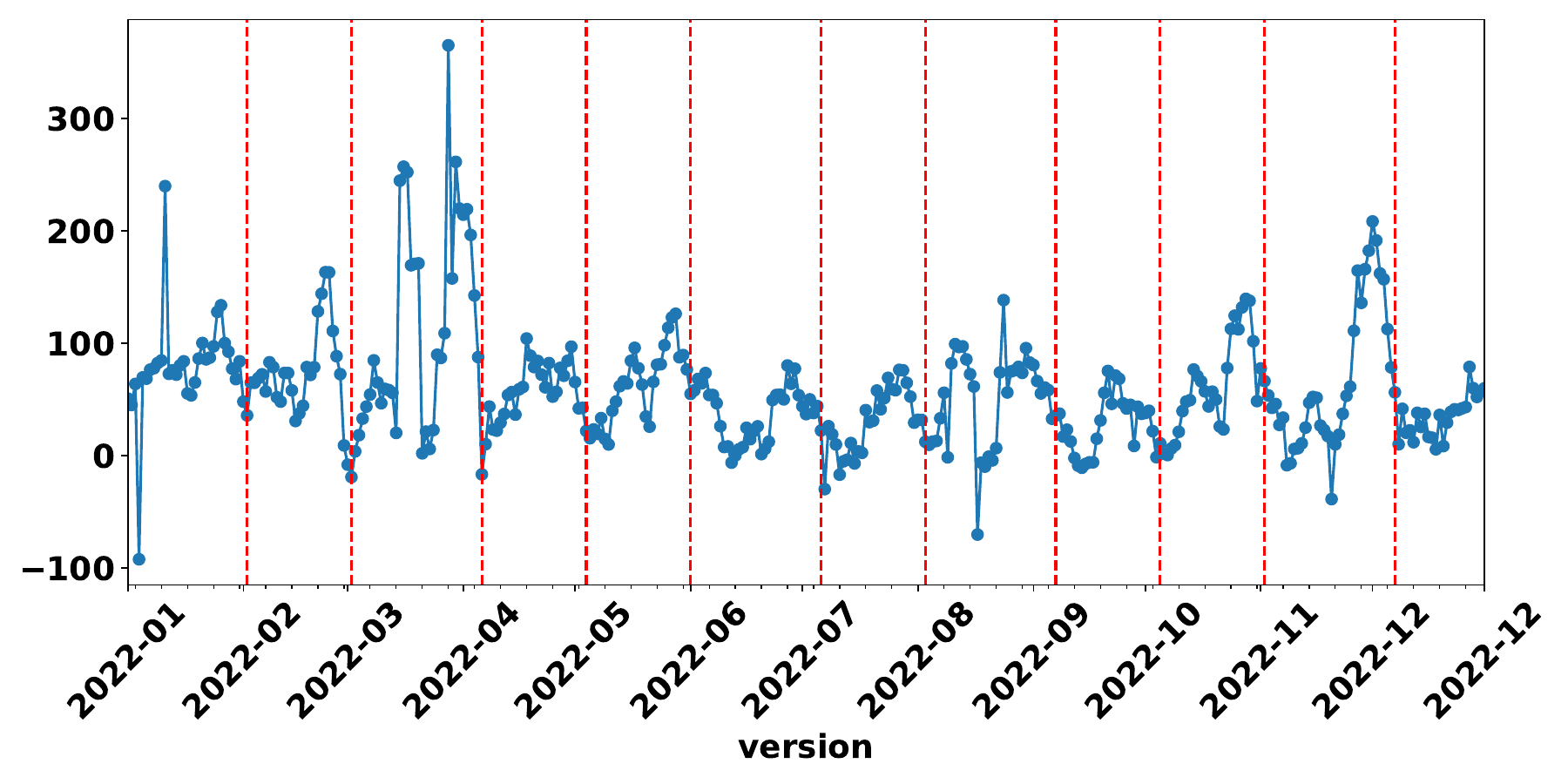}
\caption[]{Example of the SQALE Index Derivative Trends Over Time for the Home Assistant Project}
\label{fig:rq1}
\end{figure}

In this section, we reported the results of RQ1. This analysis explored the derivatives of the SQALE index trends and patterns. Figure~\ref{fig:rq1} visualizes the SQALE index derivative trends over time for the Home Assistant project throughout 2022. The red dashed vertical lines indicate the scheduled release deadlines, providing a timeline for understanding how TD accumulated in relation to these releases. Table \ref{tab:rq1} summarizes the result of our manual window coding to come up with an organized classification of the SQALE index derivative trends across the projects. The table is organized by several columns representing different trend categories, such as \textit{valley (V)} and \textit{hill (H)}, resulting from our manual window characterization process in Section \ref{Section:DataAnalysis}. The table displays the corresponding percentages of the total windows analyzed for each project. 

For Home Assistant, a significant majority of the windows were classified as \textit{valley}, accounting for 56.3\% of the windows, followed by \textit{anomalous} at 15.6\% and \textit{increase} at 12.5\%. Notably, this project has 56.3\% of the windows indicating more TD is accumulated as deadlines approach.  Similarly, the results of GitLab emphasized more on \textit{valley} patterns at 50.0\%, which also means that this project has 50.0\% of the windows indicating that more TD is accumulated as deadlines approach. Django showed a distinct preference for \textit{hill} patterns, making up 42.6\% of its total windows, with a significant number of \textit{valley} at 20.6\% and \textit{increase} at 19.1\%. The result for the Django project is characterized by intense activity mid-deadline, avoiding last-minute TD accumulation. Kubernetes displayed an overwhelming majority of \textit{hill} patterns at 74.1\%, marking it as the project with the highest concentration of single-pattern types. Unlike the Django project, the \textit{hill} pattern we observed for Kubernetes is characterized by peaks close before the deadlines. Hence, our results for Kubernetes suggested a cyclical pattern where more TD accumulated as regularly scheduled deadlines approached. Jenkins, the project with the highest total windows, presented a diverse range of patterns, with \textit{constant} patterns dominating at 43.0\%. It also showed a notable number of \textit{increase} and \textit{decrease} patterns at 17.4\% and 16.8\%, respectively, indicating significant fluctuations and stable TD accumulation over time in its dataset. A significant portion of the windows (48.0\%) in the Eclipse project were categorized as \textit{anomalous}. For the Go project, a significant proportion of the windows fell under \textit{hill} (22.2\%) and \textit{double hills} (27.8\%), similar to our results for Django. The Ansible project exhibited half of the windows in the \textit{valley} category.


\subsection{$\mathbf{RQ_2}$: The Impact of Scheduled Deadlines on the Development Commit Activities }


\begin{table*}[htbp]
\centering
\caption{Summary of RQ2 Results}
\begin{tabular}{l*{12}{r}}
\toprule
\textbf{Projects} & \multicolumn{4}{c}{\textbf{Commit Frequency}} & \multicolumn{4}{c}{\textbf{Code Churn}} & \multicolumn{4}{c}{\textbf{Files Changed}} \\
 & \textbf{Ex} & \textbf{Bf} & \textbf{Af} & \textbf{Ot} & \textbf{Ex} & \textbf{Bf} & \textbf{Af} & \textbf{Ot} & \textbf{Ex} & \textbf{Bf} & \textbf{Af} & \textbf{Ot} \\
\midrule
HomeAsst & 28\% & \textbf{53\%} & 19\% & & 9\% & 16\% & 9\% & \textbf{66\%} & \textbf{100\%} & & & \\
GitLab & 5\% & \textbf{63\%} & 16\% & 16\% & 5\% & 29\% & 7\% & \textbf{59\%} & 2\% & \textbf{50\%} & 23\% & 25\% \\
Django & 26\% & 26\% & \textbf{48\%} & & 6\% & 10\% & 16\% & \textbf{68\%} & 20\% & 22\% & 9\% & \textbf{49\%} \\
Kubernetes & 7\% & \textbf{48\%} & 30\% & 15\% & 4\% & 30\% & 18\% & \textbf{48\%} & 4\% & 26\% & 15\% & \textbf{55\%} \\
Jenkins & 13\% & 19\% & 6\% & \textbf{62\%} & 4\% & 16\% & 20\% & \textbf{60\%} & 30\% & 24\% & 8\% & \textbf{38\%} \\
Eclipse & 0\% & 38\% & 4\% & \textbf{58\%} & 0\% & 0\% & 17\% & \textbf{83\%} & 8\% & 12\% & 17\% & \textbf{63\%} \\
Go & 5\% & \textbf{39\%} & \textbf{39\%} & 17\% & 0\% & 17\% & 22\% & \textbf{61\%} & 11\% & 22\% & \textbf{39\%} & 28\% \\
Ansible &  &  &  & \textbf{100\%} &  &  &  &  \textbf{100\%} &  &  & &  \textbf{100\%} \\
\bottomrule
\end{tabular}
\label{tab:rq2}
\end{table*}

\begin{figure}[!ht]
\centering
\includegraphics[width=8.5cm, height=4cm]{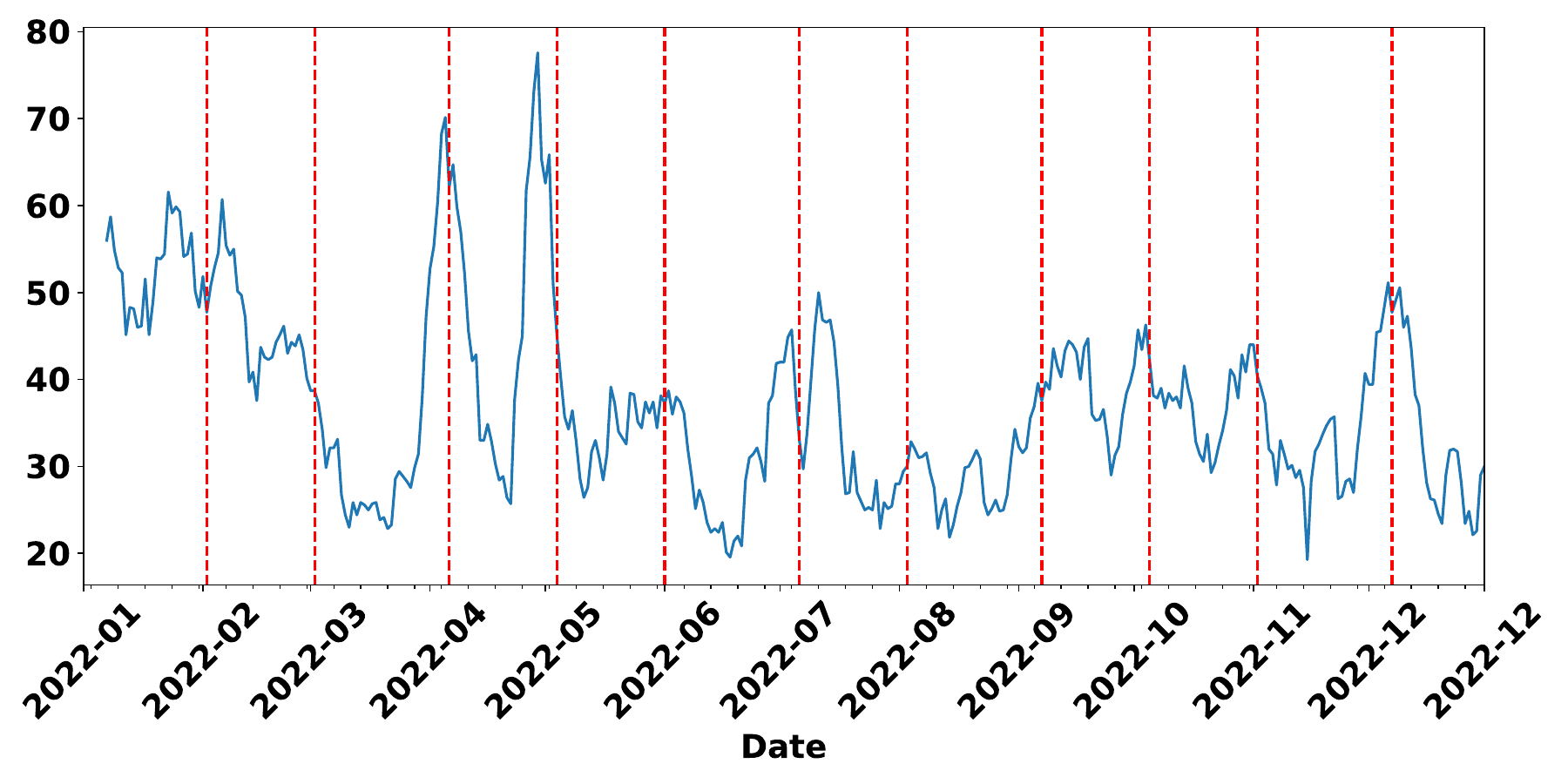}
\caption[]{Example of the Commit Frequency Trends Over Time for the Home Assistant Project}
\label{fig:rq2}
\end{figure}

In this section, we described the results of RQ2. RQ2 aims to investigate the impact of scheduled deadlines on development commit activities, focusing on daily metrics such as commit frequency, code churn, and files changed in relation to impending project deadlines. The three daily metrics were analyzed during different periods around the deadlines to understand how development activities adjusted under scheduled deadlines. Figure \ref{fig:rq2} visualizes the commit frequency trends over time for the Home Assistant project throughout 2022. The red dashed vertical lines indicate the scheduled release deadlines, providing a timeline for understanding how commit frequency accumulates in relation to the deadlines. Table \ref{tab:rq2} summarizes the results of RQ2. 

Home Assistant showed a significant portion of deadlines having commit frequency spikes occurring right before deadlines (53\%), with all of the release deadlines (100\%) having files changed occurring exactly on deadlines. The GitLab data illustrated a heavy concentration of deadlines having commit frequency (63\%) and files changed (50\%) spikes right before deadlines. Django presented a more balanced distribution with a slight emphasis on activities right after deadlines (48\% of deadlines in commit frequency). Kubernetes and Jenkins showed diverse patterns of the daily metrics in relation to deadlines. Kubernetes demonstrated a relatively even distribution across phases, whereas Jenkins showed a significant portion of activity outside the scope of deadlines (62\% in commit frequency and 60\% in code churn). Similarly, most activities in Eclipse occured outside the scope of the deadlines. Go exhibited a balanced distribution in commit frequency right before and after the deadlines (both 39\%). Ansible was unique in that all activities were exclusively categorized in \textit{Ot}, reflecting 100\% across all metrics.


\subsection{$\mathbf{RQ_3}$: The Impact of Scheduled Deadlines on the Project Issues}

\begin{table}[htbp]
\caption{Summary of RQ3 Results (Issue Creation and Closed Dates)}
\centering
\resizebox{0.47\textwidth}{!}{%
\begin{tabular}{l r r r r r r r r}
\hline
\multicolumn{1}{c}{\textbf{Projects}} & \multicolumn{4}{c}{\textbf{Issue Creation Dates}} & \multicolumn{4}{c}{\textbf{Issue Closed Dates}} \\
 & \textbf{Ex} & \textbf{Bf} & \textbf{Af} & \textbf{Ot} & \textbf{Ex} & \textbf{Bf} & \textbf{Af} & \textbf{Ot} \\
\hline
HomeAsst &  & 6\% & \textbf{85\%} & 9\% &  & \textbf{75\%} & 22\% & 3\% \\
GitLab & 2\% & \textbf{50\%} & 9\% & 39\% & 5\% & \textbf{68\%} & 18\% & 9\% \\
Django & 3\% & 37\% & \textbf{40\%} & 20\% & - & - & - & - \\
Kubernetes &  & 33\% & \textbf{56\%} & 11\% &  & 44\% & \textbf{56\%} &  \\
Jenkins & 19\% & 24\% & \textbf{34\%} & 23\% & 10\% & 11\% & 11\% & \textbf{68\%} \\
Eclipse & 8\% & 17\% & 0\% & \textbf{75\%} & 0\% & 33\% & 0\% & \textbf{67\%} \\
Go & 0\% & 0\% & 32\% & \textbf{68\%} & 0\% & 5\% & 32\% & \textbf{63\%} \\
Ansible & 0\% & 14\% & 0\% & \textbf{86\%} & 0\% & 14\% & 0\% & \textbf{86\%} \\
\hline
\end{tabular}%
}
\label{table:rq3_date}
\end{table}

\begin{table}[htbp]
\caption{Summary of RQ3 Results (Issue Labels)}
\centering
\begin{tabular}{l l r r r r}
\hline
\textbf{Projects} & \textbf{Labels} & \textbf{Ex} & \textbf{Bf} & \textbf{Af} & \textbf{Ot} \\
\hline
HomeAsst & bugfix &  & 31\% & \textbf{69\%} &  \\
     & code-quality & 3\% & \textbf{53\%} & 28\% & 16\% \\
              & breaking-change & 9\% & \textbf{47\%} & 35\% & 9\% \\
\hline
GitLab        & type::feature &  & \textbf{77\%} & 23\% &  \\
              & type::bug & 4\% & \textbf{66\%} & 30\% &  \\
              & type::maintenance & 2\% & \textbf{52\%} & 30\% & 16\% \\
\hline
Django        & Bug & 1\% & 34\% & 31\% & \textbf{34\%} \\
              & Cleanup/optimization & 9\% & 26\% & 28\% & \textbf{37\%} \\
\hline
Kubernetes    & kind/bug & 4\% & \textbf{63\%} & 26\% & 7\% \\
              & kind/cleanup &  & 26\% & \textbf{59\%} & 15\% \\
\hline
Jenkins       & Bug &  &  &  & \textbf{100\%} \\

\hline
Eclipse       & bug &  &  &  & \textbf{100\%} \\
              & help wanted &  &  &  & \textbf{100\%} \\
\hline
Go            & NeedsInvestigation & \textbf{5\%} & 16\% & 11\% & \textbf{68\%} \\
              & NeedsFix & 0\% & 5\% & 32\% & \textbf{63\%} \\
              & help wanted &  &  &  & \textbf{100\%} \\
\hline
Ansible       & bug & 14\% & 29\% & 0\% & \textbf{57\%} \\
              & needs\_revision & 0\% & \textbf{86\%} & 0\% & 14\% \\
              & has\_issue & 0\% & 14\% & 29\% & \textbf{57\%} \\
\hline
\end{tabular}
\label{table:rq3_labels}
\end{table}

\begin{figure}[!ht]
\centering
\includegraphics[width=8.5cm, height=4cm]{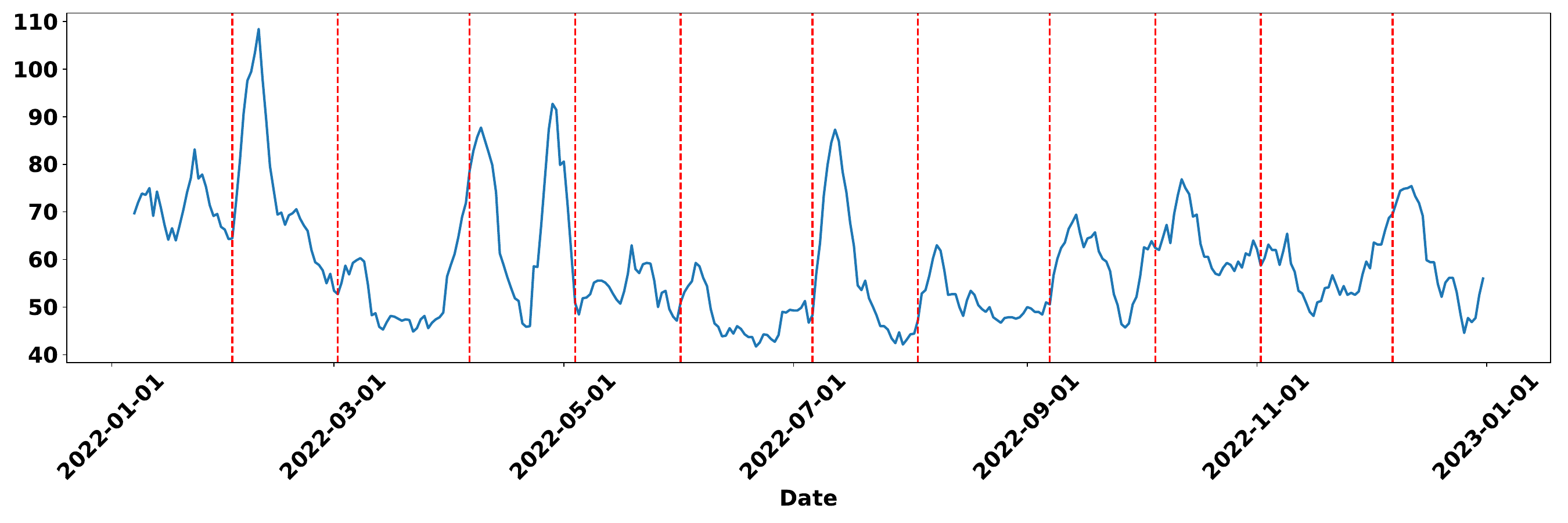}
\caption[]{Example of the Issue Creation Trends Over Time for the Home Assistant Project}
\label{fig:rq3}
\end{figure}

Here, we described the results of RQ3, summarized in Tables \ref{table:rq3_date} and \ref{table:rq3_labels}. RQ3 focuses on understanding the impact of scheduled deadlines on project issues, specifically analyzing issue tracker data regarding the creation, closure dates, and labels of issues around project deadlines. This analysis involved collecting daily metrics of issue tracker metadata and examining trends in relation to project deadlines. Figure \ref{fig:rq3} visualizes the issue creation trends over time for the Home Assistant project throughout 2022. The red dashed vertical lines indicate the scheduled release deadlines, providing a timeline for understanding how issue creation trend accumulates in relation to these releases. However, it is essential to note the limitations encountered during our analysis. Specifically, due to the unavailability of data regarding the closure dates of issues within the Django project, we could not present results for the trend of issue closure for Django. 

Across several projects, a significant number of deadlines presented a spike in issue creation right after the deadlines. Home Assistant and Kubernetes showed that a substantial proportion (85\% and 56\%, respectively) of the deadlines have spikes in issues created right after deadlines. The closure of issues also reflects interesting patterns, particularly in Home Assistant and GitLab, where 75\% and 68\%, respectively, of the deadlines have spikes right before the deadlines. However, Kubernetes exhibited a balanced closure rate before and after deadlines. For Jenkins, there were spikes in issue creation right after the deadlines (34\% of deadlines). For Eclipse, Go, and Ansible, issue creation and closure predominantly occurred not close to the deadline proximity (67–86\% of deadlines), with minimal or no activity observed exactly on, before, or after deadlines, indicating that their issue creation and closure trends were not closely aligned with the deadline proximity.

Table \ref{table:rq3_labels} further depicts the results of RQ3, focusing on the impact of scheduled deadlines on project issues, specifically through the lens of issue labels. Issue labels provide insights into the nature of issues (e.g., bug fixes, feature requests, and maintenance tasks) and how their management is influenced by project deadlines. In the analysis of RQ3, we specifically selected issue labels that describe the nature of the tasks, such as those related to bugs or maintenance. This targeted selection was used to provide a clearer understanding of how different types of tasks were influenced by approaching deadlines. Each row in the table represents data for a different project, categorizing issues by their labels, and we observed whether there were spikes \textit{Exactly on the Deadlines (Ex), Right Before (Bf), Right After (Af),} or \textit{No Spikes (Ot)}.

For Home Assistant, the labels analyzed were \textit{bugfix, code-quality, and breaking-change}. A notable trend is that a significant portion of deadlines (69\%) have spiked \textit{bugfix} issues reported right after the deadlines. In contrast, issues labeled as \textit{code-quality} and \textit{breaking-change} showed a balanced distribution, with a considerable number addressed right before the deadlines. GitLab results strongly emphasized addressing feature-related and bug-related issues right before the deadlines, with 77\% of deadlines having feature-related issues (\textit{type::feature}), 66\% of deadlines having bug-related issues (\textit{type::bug}), and 52\% of deadlines having maintenance-related issues (\textit{type::maintenance}) being created right before deadlines. Django issue creation patterns, as seen through the labels \textit{Bug} and \textit{Cleanup/optimization}, indicated a fairly even distribution of issue creation across different timeframes relative to deadlines. In Kubernetes, there was a notable emphasis on addressing bug-related issues right before the deadlines (63\%). However, cleanup-related issues showed a majority (59\% of deadlines) being handled right after the deadlines. For Jenkins, Eclipse, and Go, most data fell in the \textit{Ot} category (63–100\%), indicating that the creation of those labels was not close to the deadlines. In Ansible, labeled issues creation was also predominantly not close to deadlines, with limited creation activity spiking right before the deadlines for the  \textit{needs\_revision} label.


\section{Discussion}
\label{sec:discussion}


\textbf{Complex Dynamics of Scheduled Deadlines and Technical Debt.} Our study revealed no definitive patterns linking scheduled deadlines to TD accumulation across the selected OSS projects. While half of the selected OSS projects demonstrated clear increases in TD as deadlines approached, others showed no significant change. Importantly, \emph{we found no evidence of causality between scheduled deadlines and TD across all projects, suggesting the influence of confounding factors}. These factors may include scrupulous project management practices, project complexity, team size, resource availability, and individual developer practices, such as postponing the resolution of To-Do items or failing tests for later sessions. Such elements could heavily impact working patterns, making it difficult to draw a direct correlation between TD and scheduled deadlines. Our results indicated a potential correlation rather than causation, leaving room for potential unaccounted influences. Future research should investigate more direct or composite measures of scheduled deadlines and their relationship with TD in software development environments to understand this complex dynamic better.

\textbf{Technical Debt Accumulates as Scheduled Deadlines Approach in Some Projects.} For the Home Assistant, GitLab, and Ansible projects, we observed a majority of the windows being classified as \textit{Valley} in relation to TD accumulation. For the Kubernetes project, it exhibited a significant tendency towards \textit{Hill} patterns, with 74.1\% of the analyzed windows showing this trend as deadlines approached. Unlike the Django project, the \textit{Hill} pattern we observed for Kubernetes is characterized by peaks close to the deadline. Hence, our results suggested a \emph{cyclical pattern where more TD is accumulated as regularly scheduled deadlines approach for some projects}. This result can be attributed to the fact that continuously keeping software production running puts immediate pressure on developers~\cite{beck2000extreme}. Under deadline pressure, developers take shortcuts when developing software, making decisions to stay on schedule, which may not be in the best interests of the project quality and whose potential adverse consequences are often not fully understood~\cite{austin2001effects}.

\textbf{Commit Frequency Accumulate Near Deadlines in Some Projects.} Our results revealed that \emph{a significant portion of deadlines have commit frequency spikes right before deadlines}. Toward the end of a software development cycle, there is a natural surge in development activities to finalize features or fix bugs that must be completed for the release~\cite{ortega2010software}. The spike in commit frequency prior to deadlines corresponds with the integration phase, where developers merge branches and ensure that their code works with the main branch codebase. However, commit frequency spikes right before deadlines indicate that developers deal with increasingly intensive and challenging work at the last minute, closer to deadlines~\cite{camuto2021suite}. Dealing with such intensive and challenging tasks at the last minute makes developers incur suboptimal software practices~\cite{bavota2016large, flisar2019identification}. This result explains more TD accumulated as deadlines approached.

\textbf{Bug-related Issues Accumulate Near Deadlines in Some Projects.} Reporting and resolving bugs are key activities in software development~\cite{perez2021bug}. Our study revealed that \emph{some deadlines have spikes in bug-related issues right before deadlines}. Modern software development is a cooperative effort by self-managing, multidisciplinary teams~\cite{grassl2023exposing, hilderbrand2020engineering, storey2020software}. Release deadlines prompt the integration of various features or components that multiple developers have developed in parallel. The integration process can expose compatibility issues, leading to a spike in reported bug-related issues as different components interact within the unified system. The pre-deadline spikes in bug-related issues demonstrate a proactive push to resolve critical bugs for the release~\cite{adams2016modern}. However, developers introduce suboptimal software development practices when making last-minute fixes~\cite{bavota2016large, flisar2019identification}, contributing to TD accumulation.

\textbf{Scheduled Deadlines Have No Significant TD Accumulation Impact in Mature Projects.} Our findings revealed interesting insights into the impact of scheduled deadlines on TD accumulation, particularly in mature software projects. Contrary to the assumption that time constraints lead to rushed decisions and more TD accumulation~\cite{lenarduzzi2019empirical, guo2011tracking}, our analyses of Django, Jenkins, Eclipse, and Go suggested a different narrative. We observed no clear pattern of increased TD accumulation as deadlines approached in these projects. From Table \ref{datacollectiontable}, it is evident that all four projects are over 200 months in age, making them considerably older than the other projects analyzed in this study. The study by Molnar et al.~\cite{molnar2020long} showed that as software projects mature, they no longer introduce as much debt. Additionally, Digkas et al.~\cite{digkas2017evolution} stated that development teams increasingly emphasize repaying accumulated TD, maintaining and enhancing code quality as software projects mature. This culture is especially true for OSS projects that are recognized and of high quality within the open-source community. Hence, \emph{scheduled deadlines do not significantly impact TD accumulation in mature software projects}.

\textbf{The Impact of Team Size on TD Accumulation with Upcoming  Deadlines} We observed that \textit{TD tends to accumulate as deadlines approached in software projects with more developers}  (i.e., Home Assistant, GitLab, Kubernetes, and Ansible). As the number of contributors grows, so does the complexity of communication, integration, and collaboration~\cite{blackburn2006brooks}. Previous work has also shown that larger software teams produce software with more defects and higher complexity~\cite{bodaragama2023exploring}.

\textbf{The Impact of Deadline Timelines on TD Accumulation} Our analysis revealed that \textit{the relationship between scheduled deadlines and TD accumulation is inconsistent across projects, even among those with similar release schedules}. For instance, projects like Home Assistant and GitLab, which follow a monthly release schedule, exhibited clear patterns of increased TD accumulation as deadlines approached. In contrast, Django, which also follows a monthly schedule, did not display such trends. These inconsistencies suggest that TD accumulation dynamics may be shaped by project-specific practices rather than by the frequency of scheduled releases~\cite{codabux2017empirical}, highlighting the importance of contextual factors in understanding the impact of deadlines on code quality.

\section{Implications}
\label{sec:implications}

\textbf{For Researchers.} Our study observes an increased trend of TD accumulation as the deadlines approach, marked by commit frequency and bug-related issues spiking prior to deadlines. Accumulation of TD is due to a last-minute deal with increasingly intensive and challenging work, as well as a last-minute proactive push to resolve critical bugs for the release. This finding suggests a need to study which effective development strategies within the industry can mitigate last-minute coding rushes and the associated quality compromises. Researchers should focus on developing strategies to prevent developers from working on tasks at the last minute. Effective strategies from such research could guide teams on better practices, helping maintain code quality even under deadlines.

\textbf{For Practitioners.} Our findings indicate that while mature projects are less affected by scheduled deadlines, younger projects struggle with TD accumulation under scheduled deadlines. This disparity suggests that the industry needs to better support less mature projects by adopting practices that have proven effective in more established projects, such as increasingly emphasizing the repayment accumulated TD and maintaining code quality. Practitioners should look into established frameworks, such as the framework by Wiese et al.~\cite{wiese2021preventing}, for better project management practices that avoid end-of-cycle rushes and provide education on TD management to all team members. Moreover, creating a culture that values early and continuous attention to quality and regular debt repayment could help mitigate the impact of scheduled deadlines on project quality. Practitioners could also benefit from tools that provide real-time feedback on code quality to prevent TD accumulation.

\section{Threats to Validity}
\label{sec:threats}
In line with the classification system proposed by Runeson et al.~\cite{threats_to_validity} in their work on validity threats, we examined various elements that pose potential risks to this research. Additionally, we outlined the appropriate measures we implemented to mitigate these risks.

\textbf{Construct Validity} concerns the suitability of our operational measures in addressing our research questions. In our study, we relied solely on SonarQube to detect TD accumulated by the developers. Our study's validity depends solely on the accuracy of code debt measured by SonarQube. Any errors or biases, such as false positives in detecting code debt, in the SonarQube analysis could affect the study findings. To address this concern, we only selected the tool that, to the best of our knowledge, is the most extensively used in academia and industry for calculating TD over any other tool. This choice is based on the premise that a widely adopted and recognized tool would provide more reliable and consistent measurements of TD, thereby helping to ensure that our study accurately reflects its aim. Our methodology to analyze the last commit of each day for the SonarQube analyses presents a tradeoff between construct validity and the feasibility of the study. While this approach assumes that the last commit of a day is representative of the overall daily development efforts and codebase quality, we deemed this strategy necessary given the timespan of several development years per project. The last commit of the day is a logical point to assess the codebase since it includes all the changes finalized by the end of the working day. By focusing only on the last commit of each day, our study also establishes a consistent temporal marker for data collection. Moreover, given the observed fluctuations in TD across the projects, this design choice likely did not significantly impact the results. A potential threat to the validity of our study is analyzing SQALE index derivatives without normalizing them by size-related metrics, such as LOC. This was a deliberate choice that we consciously made in the research design phase. The SQALE index naturally increases in larger projects. Without normalization, we risked misjudging the true impact of scheduled deadlines on TD accumulation. However, according to Graf-Vlachy and Wagner~\cite{graf2023type}, ratios could be problematic when used as dependent variables~\cite{certo2020divided}. Therefore, unlike earlier studies, they chose not to calculate TD measures by dividing by the difference in LOC. A potential threat lies in the manual categorization performed by the authors. While the two authors reviewed the categorization collaboratively, there is still a possibility of subjective bias. Future work could consider defining clear thresholds for categorization to improve the manual categorization process. Lastly, we acknowledge the potential single-metric bias of relying solely on the SQALE index for TD evaluation. We chose the SQALE index metric, as it is the most commonly used and validated metric to quantify TD\footnote{https://docs.sonarsource.com/sonarqube-server/10.4/user-guide/metric-definitions/}. Prior works have utilized the SQALE index as a primary measure to quantify TD~\cite{robredo2024comparing, siavvas2019empirical}. This metric makes it particularly suitable for our longitudinal analysis of TD accumulation around scheduled deadlines, allowing us to capture the rate of TD accumulation over time. We did not account for other metrics, such as the number or severity of individual code smells, as they were outside the scope of this study. Future work could incorporate more metrics to provide a more comprehensive understanding of software quality issues under scheduled deadlines.

\textbf{Internal Validity} pertains to the degree to which the observed outcomes can be attributed to the ``treatment" rather than to other variables. In our research, as Alfayez et al.~\cite{alfayez2018exploratory} suggested, we avoided manual techniques and subjective assessments for determining TD to minimize bias and errors. Instead, we depended on SonarQube to measure TD, ensuring that all the systems we studied were evaluated under a consistent set of criteria. Another threat is the unavailability of closure date data for issues within the Django project, which prevented us from analyzing trends related to the timing of issue resolutions for Django, which could have provided further insights into how scheduled deadlines impact the resolution practices within this project. Another primary concern is the potential confounding factors influencing the perceived deadline pressure. These factors include but are not limited to the other contributor's work activities, the project complexity, the team size, resource availability, and individual developer practices and preferences, such as leaving to-do items or failing tests to be addressed in subsequent work sessions. These elements could affect developers' working patterns, complicating the association between TD and scheduled deadlines. 


\textbf{External Validity} concerns the level to which our findings can be generalized. One limitation of our study is the sample size. We only used data from eight OSS projects for our analysis. Our results and conclusions were drawn only from examining these projects, potentially affecting the generalizability of our findings. Although there is a potential for confirmation bias, the collected results provide an initial empirical demonstration of what could intuitively be expected. Expanding the sample size and variety would allow for more robust conclusions across different project environments. Another limitation is our exclusive focus on OSS systems, which limits the generalizability of our findings to closed-source or commercial projects. However, this study represents a preliminary exploration of a new topic. We focused on large OSS projects with years of active development. The scale and complexity of the selected projects provided valuable insights that could serve as a foundation for further research. Moreover, we selected OSS projects supported by actual companies that are popular within their respective domains.

\section{Conclusion And Future Work}
\label{sec:conclusions}
This study investigated the impact of scheduled deadlines on TD accumulation, analyzing the eight OSS projects through TD, commit activities, and issue tracker systems analyses. Key findings highlighted varying trends in TD accumulation across projects, with significant spikes in commit activities and specific issue reporting around release deadlines.

Future research should expand the variety of projects analyzed, including closed-source projects with strict deadlines. Examining data from closed-source projects would offer a more complete insight into the connection between TD accumulation and scheduled deadlines. Analyzing more TD metrics is another critical area, as relying solely on the SQALE index limits the scope of analysis. Lastly, a particularly promising area of research involves developer profiling to understand how individual behaviors and coding practices contribute to TD accumulation, which could lead to a more personalized understanding of TD accumulation. Understanding these individual developer behaviors would provide more practical insights and inform better management practices. 

\begin{acks}
This study is partly supported by the Natural Sciences and Engineering Research Council of Canada (NSERC), RGPIN-2021-04232 and DGECR-2021-00283, and an NSERC Collaborative Research and Training Experience (CREATE) grant on Software Analytics at the University of Saskatchewan.
\end{acks}

\balance
\bibliographystyle{ACM-Reference-Format}
\bibliography{bibfile}
\end{document}